\documentclass[10pt]{article}

\usepackage{booktabs}
\usepackage{tikz}
\usepackage{verbatim}
\usepackage{subfigure}
\usepackage[english]{babel}
\usepackage{multirow}
\usepackage{makecell}

\usepackage{amsmath}
\usepackage{amssymb}
\usepackage{graphicx}
\usepackage{cite}
\usepackage{color}

\usepackage[labelfont=bf,labelsep=period,justification=raggedright]{caption}

\makeatletter
\renewcommand{\@biblabel}[1]{\quad#1.}
\makeatother

\date{}

\begin{document}

\begin{flushleft}
{\Large
\textbf{Cumulative Effect in Information Diffusion: A Comprehensive Empirical Study on Microblogging Network}
}
\\
Peng Bao,
Hua-Wei Shen$^{\ast}$,
Wei Chen,
Xue-Qi Cheng
\\
\bf{Institute of Computing Technology, Chinese Academy of Sciences, Beijing, China}
\\
$\ast$ E-mail: shenhuawei@ict.ac.cn
\end{flushleft}

\section*{Abstract}
Cumulative effect in social contagions underlies many studies on the spread of innovation, behaviors, and influence. However, few large-scale empirical studies are conducted to validate the existence of cumulative effect in the information diffusion on social networks. In this paper, using the population-scale dataset from the largest Chinese microblogging website, we conduct a comprehensive study on the cumulative effect in information diffusion. We base our study on the diffusion network of each message, where nodes are the involved users and links are the following relationships among them. We find that multiple exposures to the same message indeed increase the possibility of forwarding it. However, additional exposures cannot further improve the chance of forwarding when the number of exposures crosses its peak at two. This finding questions the cumulative effect hypothesis in information diffusion. Furthermore, to clarify the forwarding preference among users, we investigate both the structural motif of the diffusion network and the temporal pattern of information diffusion process among users. The patterns provide vital insight for understanding the variation of message popularity and explain the characteristics of diffusion networks.

\section*{Introduction}

We are witnessing the emergence and rapid proliferation of various social applications, including resource sharing sites (e.g., Flickr, Youtube), blogs (e.g., Bloggers, LiveJournal), social networks (e.g., Facebook, Myspace), and microblogs (e.g., Twitter, Sina Weibo). These social applications facilitate users to produce, share, and consume online content. A prominent characteristic of these systems is the relationships formed among users. These relationships can be described by networks, where nodes represent users and links denote the relations or interactions among users. Many efforts have been made to understand the structure of theses networks\cite{Newman2010}. Recently, much research attention is paid to various dynamics on these networks, investigating users' tendency to engage in activities such as forwarding messages, linking to articles, joining groups, purchasing products, or becoming fans of certain pages after their friends have done\cite{Adar2004, Backstrom2006, Cosley2010, Crane2008, Gruhl2004, Leskovec2007a, Leskovec2007b, Liben-Nowell2008, Sun2009}.

Existing studies mainly focused on identifying the properties of these dynamics and the potential principles governing them\cite{Rogers1995, Strang1998, Lazarsfeld1968}. Scientists have noticed several salient phenomena about information diffusion on networks and the evolution of underlying networks, including the rich-get-richer phenomenon\cite{Barabasi1999}, burst\cite{Barabasi2005}, the stability constrains\cite{Perotti2009}, homophily\cite{McPherson2001}, clustering\cite{Girvan2002}, structural balance\cite{Marvel2009}, and two-step flow\cite{Katz1957}. However, it is still an open problem to understand the mechanisms of information diffusion on networks. Whether there exists the ``cumulative effect'', i.e., multiple exposures to a message can increase the possibility of forwarding it? Are there fundamental differences among the mechanisms underlying the diffusion of various messages? Does the relevant topic or the associated event of messages help explain the distinct characteristics of these messages? More importantly, are there some structural or temporal patterns frequently occurring in the processes of information diffusion?

With the increasing availability of data recording the information diffusion on social networks, many efforts have been made to study the effect of multiple exposures on social networks. Using the data from LiveJournal and DBLP, Backstrom et al. found that the propensity of individuals to join communities was dominated by a ``diminishing return'' property\cite{Backstrom2006}. Leskovec et al. examined the probability of purchasing a product as a function of the number of received recommendations about the product\cite{Leskovec2007a}. They observed a saturation point after receiving around $10$ recommendations. Romero et al. studied the mechanics of information diffusion by comparing the information diffusion process across different topics on Twitter\cite{Romero2011}. They found that the effect of multiple exposures decays rapidly for hashtags representing idioms and neologisms. Ugander et al. found that the probability of contagion was tightly controlled by the number of connected components in an individual's neighborhood, rather than by the actual size of neighborhood\cite{Ugander2012}. In addition, Milo et al. defined ``network motifs" and found them in networks from biochemistry, neurobiology, ecology, and engineering\cite{Milo2002}. Zhang et al. proposed a new mechanism for the local organization and tested potential theory\cite{Zhang2013}. They found that the Bi-fan structure was the most favored local structure in directed networks. However, recent works mainly focused on the diffusion of innovation, the adoption of new products, and the spread of certain behaviors. It is still unclear whether these findings are applicable to the information diffusion on microblogging networks.

In this paper, to understand the mechanism of information diffusion on social networks, we conduct a comprehensive empirical analysis on a population-scale dataset from Sina Weibo, the largest Chinese microblogging website. We study the statistics of diffusion network which characterizes the relationship among the individuals involved in diffusion process. We then investigate the cumulative effect of multiple exposures during the spread process of messages, with or without URLs and events. We find a peak in the curve of forwarding probability at 2 exposures and a subsequent slow drop. We also find that the probability of forwarding messages with URL or events are significantly higher than the other messages. When examining the exposure curves corresponding to different events, we find that the exposure curve is heavily affected by outside intervention, such as restrictions on media coverage. Furthermore, we investigate the structural and temporal patterns frequently occurring in information diffusion. These findings provide us great insights in understanding the fundamental mechanism of information diffusion and predicting the forwarding behavior of individuals.

\section*{Results}

\subsection*{Diffusion network}

To study the information diffusion on social networks, we represent the cascade of each message as a diffusion network. For each message, its diffusion network is a directed network where each node is a user who involves in the diffusion of this message. A link from user $u$ to user $v$ denotes that $v$ receives the message from $u$ and then forwards it. To be sure, one user can forward a message more than one time. In this paper, when constructing the diffusion network of a message, we only consider one user's first forwarding behavior of the message. In a diffusion network, there is only one node having no incoming link. We call this node the root node of diffusion network because this node corresponds to the source user of message. Similarly, we call the nodes without outgoing link as leaf nodes.

\begin{figure}[!htb]
\begin{center}
\includegraphics[height=2.5in, width=4in]{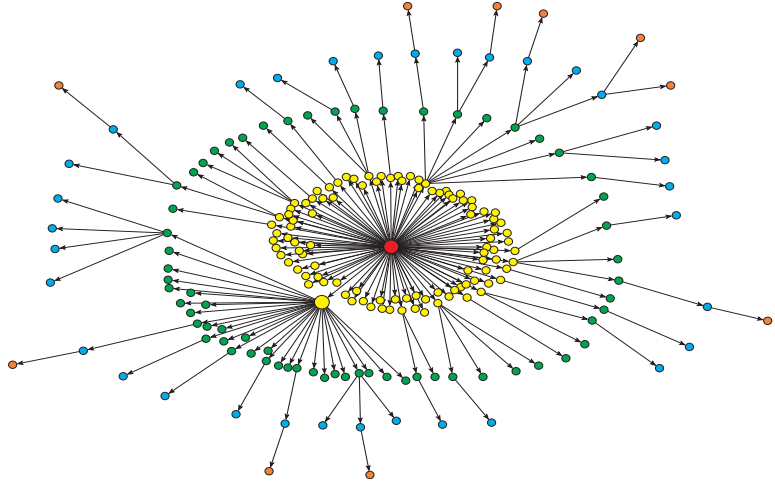}
\end{center}
\caption{
{\bf An example of diffusion network.} Colors differentiate the nodes in different layers. The root node is colored in red and its large outgoing degree indicates the early popularity of the message. The large node in yellow triggers the further spread of this message.
}
\label{fig1}
\end{figure}

Diffusion network provides us important descriptive information for the cascade of a message. On one hand, the outgoing degree of a node characterizes its amplification factor at the diffusion process of message. The nodes with larger outgoing degrees are usually the so-called ``opinion leaders''~\cite{Katz1957} and are essential to the popularity of a message. By inspecting the outgoing degree in diffusion network, we can easily identify these opinion leaders. On the other hand, each path from the root node to a leaf node depicts a forwarding trajectory of message. To a certain extent, the maximum length of all the paths reflects the penetration capability of message. Furthermore, a diffusion network generally has multiple layers. The nodes in the same layer have the same distance from the root node. Finally, the size of a diffusion network characterizes the popularity of the corresponding message. Figure~\ref{fig1} gives an example of diffusion network. The root node, colored in red, has a large outgoing degree and thus promotes the early popularity of the message. The large node in yellow is another node with a large outgoing degree, triggering a new spread range for the message.

\begin{figure}[b]
\begin{center}
\includegraphics[width=6in]{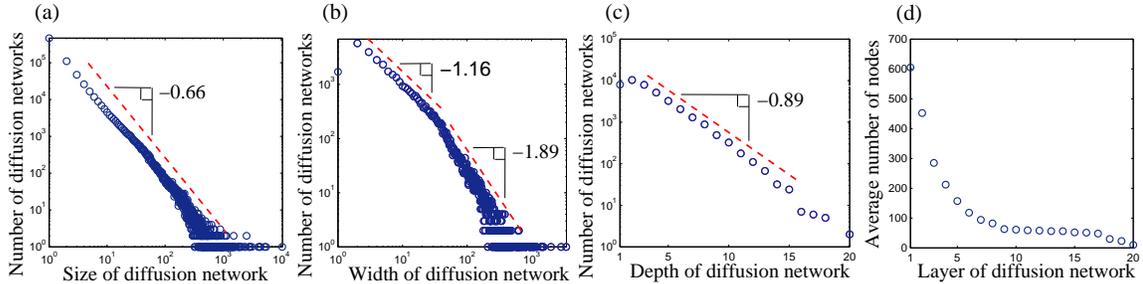}
\end{center}
\caption{
{\bf Statistics of diffusion networks.}  (a) The distribution of size of diffusion network. (b) The distribution of width of diffusion network. (c) The distribution of depth of diffusion network. (d) The average number of nodes with respect to the layer of diffusion network.
}
\label{fig2}
\end{figure}

We adopt three quantities to characterize the properties of diffusion network, i.e., the \textit{size}, \textit{depth} and \textit{width} of diffusion network. The size of a diffusion network is the number of nodes in the diffusion network and reflects the popularity of message among users. The depth of a diffusion network is the length of the longest path from the root node to leaf nodes. The width of a diffusion network is the number of nodes in the layer with the largest number of nodes. As shown in Figure~\ref{fig2}(a), the size of diffusion network follows a power law distribution with exponent 0.66, indicating that the popularity of messages is unequally distributed. This poses a big challenge for predicting the popularity of messages\cite{Yang2011,Szabo2010,Lerman2010,Hong2011}. Figure~\ref{fig2}(b) shows the distribution of width over all diffusion networks. The width distribution can be well fitted with a two-stage power law distribution with exponents respectively being $1.16$ and $1.89$. Figure~\ref{fig2}(c) shows the distribution of depth over all diffusion networks. The depth roughly follows an exponential distribution with exponent 0.89, indicating that the majority of diffusion networks have shallow depth. To characterize the shallow structure of diffusion network, we further investigate the average number of nodes in each layer of diffusion networks. As shown in Figure~\ref{fig2}(d), the average number of nodes decreases dramatically with respect to the depth of layer. The majority of nodes appear in the first $5$ layers of diffusion network.

\subsection*{Temporal characteristics of information diffusion}

Information diffusion is a dynamical process on social networks. Besides the structural characteristics depicted in the previous section, information diffusion also exhibits several temporal patterns which are the focus of this section.

\begin{figure}[htb]
\begin{center}
\includegraphics[width=5.5in]{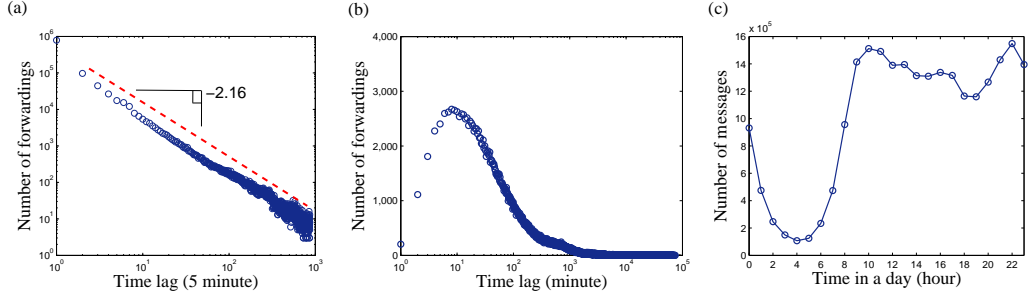}
\end{center}
\caption{
{\bf Temporal characteristics of information diffusion.} (a) The distribution of time lag between two subsequent forwarding behaviors. (b) The distribution of time lag between each forwarding behavior and the submitting behavior of message. (c) The hourly activity of users. 
}
\label{fig3}
\end{figure}

We further analyze the time lag of forwarding behaviors in diffusion process. Figure~\ref{fig3}(a) shows the distribution of time lag between two subsequent forwarding behaviors in the resolution of five minutes from the cascades of all messages, which follows a power law distribution with exponent 2.16. In addition, Figure~\ref{fig3}(b) gives the distribution of the time lag between each forwarding behavior and the submitting behavior of the corresponding message over the whole cascades. This distribution roughly follows a log-normal distribution with a peak at 10 minutes. Indeed, after a message is submitted by a user, it usually takes several minutes to be forwarded by other users, which may result from the fact that users are not always online and they check the message at a certain rate. Therefore, if a user is not active in certain period, the messages submitted will need to wait for a long time to be forwarded by this user. As a typical example, users are usually active at days and not active at nights. 

To verify the activity pattern of users, we investigate the number of messages posted hourly. Figure~\ref{fig3}(c) shows the hourly activity of users. We can see that users are active between $10$am-$10$pm and are not active between $1$am-$7$am. 

\subsection*{Cumulative effect of multiple exposures}

We now turn to the diffusion dynamics of messages on social networks. Specifically, we study the cumulative effect of multiple exposures, i.e., a user is more likely to forward a message if this user is exposed to the message for more times. There are two assumptions about the cumulative effect of multiple exposures. The first one claims that a user's multiple exposures to a message will always increase the possibility that the user forwards this message. The second one insists that more exposures will not increase the forwarding possibility if a user has ever been exposed to the message but does not forward it.

To investigate the cumulative effect of multiple exposures, we need to capture the number of exposures before a user forwards a message. For this purpose, we define that a user is \emph{k-exposed} to a message if the user has received the message for $k$ times but still does not forward it. Using the ordinal time estimate method~\cite{Romero2011}, we denote $W(k)$ the number of users who are $k$-exposed to a message at certain time, and $R(k)$ the number of users who forward the message directly after being $k$-exposed to the message. We then calculate the probability $P(k)$ that a $k$-exposed user forwards the message before this user becomes $(k+1)$-exposed, i.e., $P(k)=R(k)/W(k)$.

\begin{figure}[!htbp]
\begin{center}
\includegraphics[width=6in]{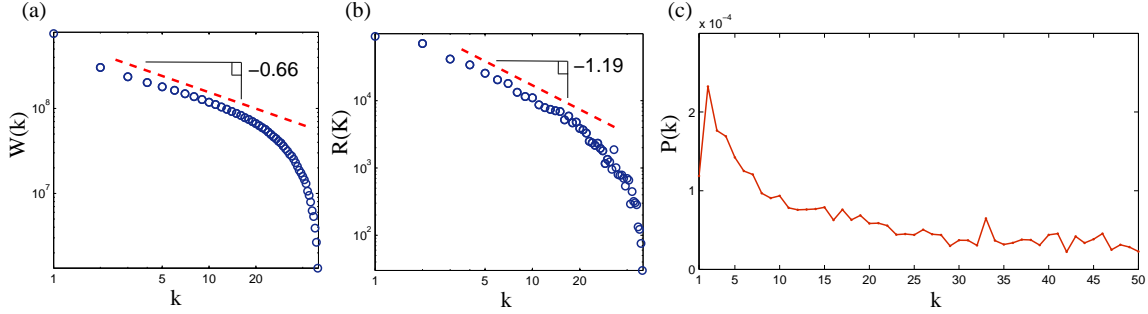}
\end{center}
\caption{
{\bf Exposure curve.}  (a) The distribution of $W(k)$, where $W(k)$ denotes the number of users who are $k$-exposed to a message at certain time. (b) The distribution of $R(k)$, where $R(k)$ denotes the number of users who forward a message directly after being $k$-exposed to it. (c) The probability of forwarding a message as a function of the number of exposures over all cascades.
}
\label{fig4}
\end{figure}

With the above definitions, we empirically study the forwarding probability $P(k)$ using all the messages forwarded by more than $10$ users. To alleviate the influence from activity pattern of users, we only consider the messages posted between $10$am and $10$pm per day, which is the active period as depicted in Figure~\ref{fig3}(c). Figure~\ref{fig4}(a) -(b) show $W(k)$ and $R(k)$ with respect to the number $k$ of exposures. Both $W(k)$ and $R(k)$ roughly follow power law distributions with cutoffs and the exponents are respectively $0.66$ and $1.19$. Figure~\ref{fig4}(c) gives the forwarding probability $P(k)$ as a function of the number $k$ of exposures. We can see that there is a peak in the curve of forwarding probability $P(k)$ at the place of $2$ exposures. After the peak, the value of $P(k)$ drops in the power law manner. These findings can provide some insights for making viral marketing strategies, such as the product promotion campaign and influence maximization\cite{Kempe2003}. In addition, the exposure curve $P(k)$ of each user can help us understand users' forwarding behavior and further identify the users that are critical to trigger a diffusion from the perspective of sender and receiver. Actually, Aral et al. have moved along this line and suggested that influential people with influential followers may be instrumental in the spread of product on social networks~\cite{Aral2012}. 

\begin{figure}[!htbp]
\begin{center}
\includegraphics[width=5.1in]{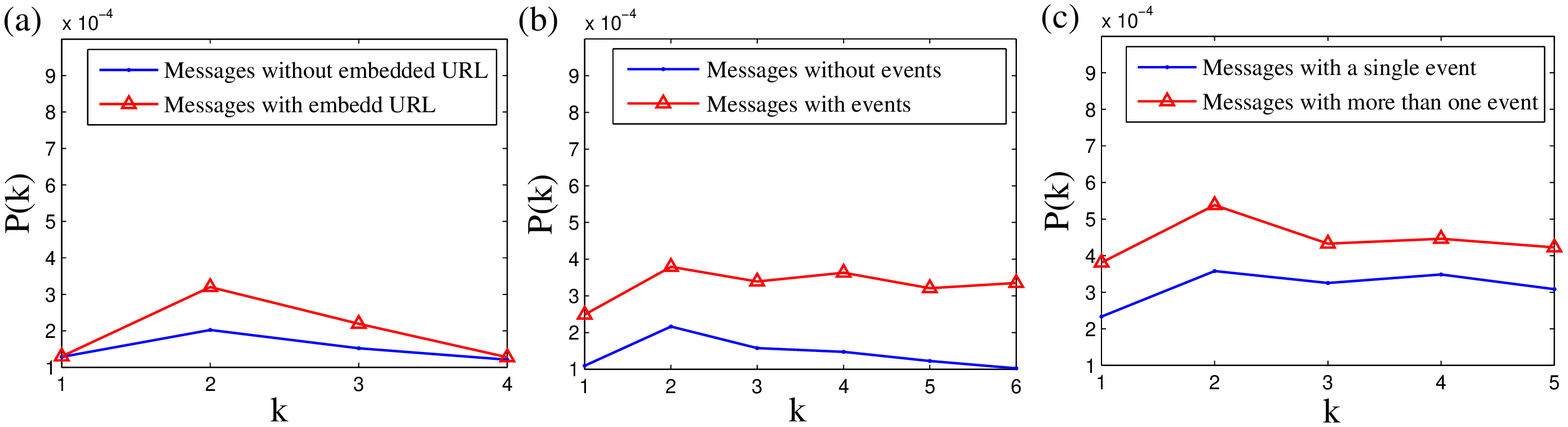}
\end{center}
\caption{
{\bf The variation of exposure curve for different kinds of messages.} (a) The probability of forwarding a message with and without embedded URL. (b) The probability of forwarding a message with and without events. (c) The probability of forwarding a message with more than one event and with a single event.
}
\label{fig5}
\end{figure}

To understand the variation of exposure curve for different messages, we classify messages into different categories and compare the exposure curves of each category. Three criteria are adopted to classify the messages into categories: (1) messages with embedded URL versus messages without embedded URL; (2) messages with events versus messages without events; and (3) messages with a single event versus messages with more than one event. The comparison of exposure curves is shown in Figure~\ref{fig5}. We can see that the probability of forwarding a message with embedded URL is higher than that of forwarding a message without embedded URL, as shown in Figure~\ref{fig5}(a). The probability of forwarding a message with events is higher than that of forwarding a message without events, as shown in Figure~\ref{fig5}(b). The probability of forwarding a message with more than one event is higher than that of forwarding a message with a single event, as shown in Figure~\ref{fig5}(c). In addition, the probability of forwarding a message with embedded URL or with events is higher than $P(k)$ over all messages which is depicted in Figure~\ref{fig4}(c). These findings indicate that users are prone to forward messages containing more information, e.g., with a URL providing additional information or with events implying much more information related to the message. In addition, a message with events can trigger more discussions about the events.

\begin{figure}[!htbp]
\begin{center}
\includegraphics[width=5in]{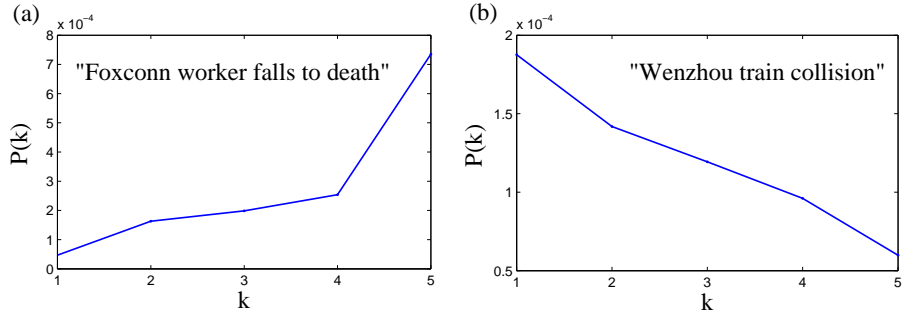}
\end{center}
\caption{
{\bf Two particular cases of exposure curve.}  (a) The exposure curve for the event ``Foxconn worker falls to death'', in which $P(k)$ increases with number of exposures. (b) The exposure curve for the event ``Wenzhou train collision'', in which $P(k)$ decreases with the number of exposures.
}
\label{fig6}
\end{figure}

We further investigate the exposure curves of messages corresponding to individual event. The majority of them are similar to the overall shape in Figure~\ref{fig4}(c). In particular, we notice that $P(k)$ increases with more exposures to a message for some examples while $P(k)$ decrease with more exposures for others. As examples, Figure 6(a) and Figure 6(b) show the exposure curves for the event ``Foxconn worker falls to death'' and ``Wenzhou train collision'' respectively. This kind of difference lies in the specific contexts of these messages. The ``Foxconn worker falls to death'' event occurred successively in a short period of time and prompted wide and in-depth discussions about laborers' working condition and payment. As a result, the more exposures one is exposed to, the higher probability one might become involved. However, the ``Wenzhou train collision'' event happened suddenly. Two high-speed trains collided with each other, 40 people were killed, and at least 192 were injured. Officials responded to the accident by hastily concluding rescue operations and ordering the burial of the derailed cars. These actions elicited strong criticism from Chinese media and online communities. In response, the government issued directives to restrict media coverage, which was met with limited compliance, even on state-owned networks. Thus, the distinct forwarding curve $P(k)$ for the messages about this event is partly caused by outside intervention\footnote{http://en.wikipedia.org/wiki/Wenzhou\_train\_collision}.

\subsection*{Analysis of structural motif and temporal pattern}
In this section, we study the structural and temporal patterns in diffusion process to answer the following question: for an individual, who is exposed to a message for multiple times from whose followees, whom will the individual echo, the first one, the last one, or the most influential one? As shown in Figure 4, among all the forwarding behaviors, 60 percent are conducted after the target user is exposed to a message for two times. Thus, we only focus on the 2-exposure case that one user is exposed to a message for two times and then forwards it.

For the 2-exposure case, without loss of generality, we assume that one exposure is from user A and the other exposure is from user B. Then, according to the relationship between A and B, we have three types of structural motifs, which are respectively ``Diverse motif'' if there is no direct relationship between A and B, ``Reciprocal motif'' if A follows B and B also follows A, and ``Unidirectional motif'' if A follows B or B follows A.

\begin{table}[!htbp]
\caption{Statistics of three types of structural motifs} 
\renewcommand{\arraystretch}{2}
\centering 
\begin{tabular}{l c c c} 
\toprule
	Types of structural motifs	
	& \multirow{1}{*}{
		\begin{tikzpicture}[node distance=10mm]
	        \tikzstyle{every node}=[fill=blue!30!white,draw=black,minimum size=6mm,circle]
	        \node (A1) {A};
	        \node (B1) [right of=A1] {B};
		\end{tikzpicture}
	} & \multirow{1}{*}{
		\begin{tikzpicture}[node distance=10mm]
	        \tikzstyle{every node}=[fill=blue!30!white,draw=black,minimum size=6mm,circle]
	        \node (A1) {A};
	        \node (B1) [right of=A1] {B};
	        \draw[->, dashed] (A1)--(B1);
		\end{tikzpicture}
	} & \multirow{1}{*}{
		\begin{tikzpicture}[node distance=10mm]
	        \tikzstyle{every node}=[fill=blue!30!white,draw=black,minimum size=6mm,circle]
	        \node (A1) {A};
	        \node (B1) [right of=A1] {B};
	        \draw[->, dashed] (A1)--(B1);
	        \draw[->, dashed] ([yshift= -1mm]B1.west)--([yshift= -1mm]A1.east);
		\end{tikzpicture}
	}  \\ 
\hline
\midrule 
	Description & ``Diverse" & ``Unidirectional" & ``Reciprocal" \\
	\% over all messages & 76.5\% & 9.0\% & 14.5\% \\
\hline
	\% over messages with embedded URL & 75.5\% & 10.3\% & 14.2\%\\
	\% over messages without embedded URL & 76.8\% & 8.6\% & 14.5\%\\
	\% over messages without events & 74.2\% & 9.8\% & 16.0\% \\
	\% over messages with events & \textbf{83.9\%} & 6.7\% & 9.4\% \\
	\% over messages with a single event & \textbf{83.1\%} & 6.6\% & 10.3\% \\
	\% over messages with more than one event & \textbf{87.8\%} & 7.0\% & 5.2\% \\
\hline
	\% over messages with popularity between 10$\sim$100 & 50.8\% & 17.0\% & 32.2\% \\
	\% over messages with popularity between 100$\sim$1000 & 76.0\% &9.6\% & 14.4\% \\
	\% over messages with popularity between 1000$\sim$10000 & 82.6\% & 7.0\% & 10.4\% \\
	\% over messages with popularity larger than 10000 & 87.8\% & 5.0\% & 7.3\% \\
\bottomrule 
\end{tabular}
\label{table1} 
\end{table}

Table~\ref{table1} shows the percentage of the three types of two-node motifs over all data set. We can see that the percentage of ``Diverse motif'' over the whole data set is 76.5\%, which is significantly higher than the other two patterns. For a detailed analysis, we further report the percentage of the three types of two-node motifs in different categories depicted in the previous section. We find that the percentage of ``Diverse motif" over messages with events is 83.9\%, which is higher than that over messages without events. The percentage of ``Diverse motif" is even higher, i.e., 87.8\%, over messages with more than one event. In addition, the percentage of ``Diverse motif" over messages with a single event is 83.1\%, which is still higher than the average percentage of that over all data set. However, the percentage of the three types of motifs over messages with or without URL is close to that over all data set. One possible explanation for these findings is that a message with events might trigger more discussions about the events, and then an individual is more likely to be exposed to the message for multiple times. 

Furthermore, we divide the messages into four different classes according to their popularity. These classes are class 0 - Messages that were forwarded by 10$\sim$100, class 1 - Messages that were forwarded by 100$\sim$1000, class 2 - Messages that were forwarded by 1000$\sim$10000, and class 3 - Messages that were forwarded more than 10000 times. As shown in Table~\ref{table1}, from class 0 to class 3, the percentage of the ``Diverse motif" increases while the other two decrease. This indicates that a diverse group of individuals spread a message to wider audience than a dense group.

\begin{table}[!htbp]
\caption{Statistics of temporal patterns} 
\renewcommand{\arraystretch}{2}
\centering 
\begin{tabular}{l c c} 
\toprule
	Types of behavior of X & X forwards from A & X forwards from B \\ 
\hline
\midrule 
	A is earlier than B & 14.5\% & 85.5\% \\
	A's indegree is bigger than B's and A is earlier than B & 38.9\% & 61.1\% \\
	A is the source of message & 43.7\% & 56.3\% \\
\bottomrule 
\end{tabular}
\label{table2} 
\end{table}

We now turn to the problem: if an individual X is exposed to a message from A and B, whom will X forward the message from? The earlier one A, the one with bigger indegree, or the source of the message? We analyze this ``forwarding whom" problem in our data set. The results are shown in Table~\ref{table2}. 
When a user is exposed for twice, the percentage of the temporal pattern that X forwards from the latter exposure is 85.5\%, while the percentage of the pattern that X forwards from the earlier one is just 14.5\%. Furthermore, if user A's indegree on social graph is bigger than B's, the percentage of the temporal pattern that X forwards from A is 38.9\%. If A is the source of message, the percentage is 43.7\%. The results on the temporal patterns in information diffusion provide several empirical evidence for understanding the forwarding behavior of individuals and the evolution of diffusion network. 


\section*{Discussion}
In this paper, we have analyzed the information diffusion on the microblogging network in the microscopic perspective. Our study is conducted on the biggest microblogging network in China. Specifically, we have studied the cumulative effect of multiple exposures on Sina Weibo. We also studied the effect on the spread of a message that was divided into groups according to the contents of each event in detail. We have observed a peak in the probability of forwarding at 2 exposures and then a slow drop. We have found that the probability of forwarding a message that containing embedded URL, a single event related, and multi-event related was significantly higher. We have examined the exposure curves corresponding to different events specifically. To our surprise, we have found that the exposure curve could be affected by outside intervention, such as restrictions on media coverage. In addition, we investigated the structural and temporal patterns that of a higher probability to appear in diffusion processes. These findings provide us great insights in understanding the fundamental mechanism of information diffusion and in predicting the  behavior of forwarding for an individual.

A long list of extensions can be conducted based on our findings. Examples include deep exploration on the relationship between the final popularity of a message and the characteristics of the networks spanned by early adopters, i.e., the users who view or forward the content in the early stage of content dissemination. We will further study the various roles played by individuals on social network. A probabilistic view might be introduced to explain the cumulative effect of multiple exposures. Besides, one is also encouraged to discover more temporal characteristics by time series analysis. As future work, we will be devoted to the modeling of forwarding behavior of individuals and the popularity prediction problem.

\section*{Materials and Methods}

The data set is collected from the most popular Chinese microblogging service, namely Sina Weibo. Sina Weibo has more than $300$ million registered users and generates about $100$ million messages per day. The length of each message is no larger than $140$ characters. Users obtain messages from other users through following relationships among them. Each following relationship is a directed link from the follower to the followee. For each user, the messages from his/her followees are ranked chronologically. Users can both deliver new messages and forward other users' messages.

We get the data set from the WISE 2012 Challenge\footnote{http://www.wise2012.cs.ucy.ac.cy/challenge.html}. This data set is crawled via the API provided by Sina Weibo. According to Sina Weibo's Terms of Services, both the user IDs and the message IDs are anonymized. The content of messages is also removed. However, some messages are annotated with events. Each event has the terms used to identify the event and a link to Wikipedia (http://wikipedia.org) page containing descriptions to the event.

In this paper, we only use the messages that was originally posted to Sina Weibo between July 1, 2011 and July 31, 2011. There are $16.6$ million messages. For each message, we collect its forwarding information between July 1, 2011 and August 31, 2011. For each forwarding of a message, the recorded information contains the anonymized users, the timestamp of this forwarding, and the forwarding path containing all the anonymized users in the path from the original user to the current user. The timestamp is in the resolution of seconds.

In addition, the data set also contains a snapshot of the social network recording the followships among users. The social network contains $58.6$ million users and $265.5$ million followships among them.

\section*{Acknowledgments}
We acknowledge Junming Huang, Suqi Cheng and Yongqing Wang for helpful discussions.


\section*{Figure Legends}
Figure 1. {\bf An example of diffusion network.} Colors differentiate the nodes in different layers. The root node is colored in red and its large outgoing degree indicates the early popularity of the message. The large node in yellow triggers the further spread of this message.

Figure 2. {\bf Statistics of diffusion networks.}  (a) The distribution of size of diffusion network. (b) The distribution of width of diffusion network. (c) The distribution of depth of diffusion network. (d) The average number of nodes with respect to the layer of diffusion network.

Figure 3. {\bf Temporal characteristics of information diffusion.} (a) The distribution of time lag between two subsequent forwarding behaviors. (b) The distribution of time lag between each forwarding behavior and the submitting behavior of message. (c) The hourly activity of users. 

Figure 4. {\bf Exposure curve.}  (a) The distribution of $W(k)$, where $W(k)$ denotes the number of users who are $k$-exposed to a message at certain time. (b) The distribution of $R(k)$, where $R(k)$ denotes the number of users who forward a message directly after being $k$-exposed to it. (c) The probability of forwarding a message as a function of the number of exposures over all cascades.

Figure 5. {\bf The variation of exposure curve for different kinds of messages.} (a) The probability of forwarding a message with and without embedded URL. (b) The probability of forwarding a message with and without events. (c) The probability of forwarding a message with more than one event and with a single event.

Figure 6. {\bf Two particular cases of exposure curve.}  (a) The exposure curve for the event ``Foxconn worker falls to death'', in which $P(k)$ increases with number of exposures. (b) The exposure curve for the event ``Wenzhou train collision'', in which $P(k)$ decreases with the number of exposures.

\section*{Tables}

\end{document}